\documentclass[namedreferences,hyperref,optionalrh,solaromanenum,showbiblabels]{spr-sola}
\usepackage{graphicx}        % For eps figures, newer & more powerfull
\usepackage{color}           % For color text: \color command
\usepackage[table]{xcolor}
\usepackage{hyperref}
\usepackage{url}
\usepackage{breakurl}  
\hypersetup{colorlinks=true,citecolor=blue}  
\chardef\us=`\_

\begin{document}
\begin{article}
\begin{frontmatter}
\title{Machine Learning–Based Characterization of Solar p‑Mode Frequency Shifts during Solar Cycle 25}
\author[addressref=aff1]{Rekha Jain\orcid{0000-0002-0080-5445}}
\author[addressref={aff2}]{Akash Kumar\orcid{0000-0003-4836-2126}}
\author[addressref={aff3}]
{\inits{S.T.}Sushanta C. Tripathy\orcid{0000-0002-4995-6180}}
%\author{\inits{}\fnm{}~\lnm{}\orcid{}}
%   NOTE:  Just one corresponding author [corref]
\address[id=aff1]{School of Mathematical and Physical Sciences, University of Sheffield, S3 7RH UK {R.Jain@sheffield.ac.uk}}
\address[id=aff2]{School of Mechanical, Aerospace and Civil Engineering, University of Sheffield, S1 3JD UK}
\address[id=aff3]{National Solar Observatory, Boulder, CO 80303,  USA}

\runningauthor{Jain, Kumar and Tripathy}
\runningtitle{\textit{Machine Learning–Based Characterization of Solar p‑Mode Frequency Shifts}}
\begin{abstract}
The solar interior is probed by the properties of the Sun’s acoustic oscillations ($p$-modes) observed on the solar surface. The frequencies of these $p$-modes measured in the last three decades show long term variation similar to the $\sim$11 year cyclic behaviour exhibited by 10.7 cm radio flux, sunspot numbers and other solar activity indices.
It is also now established that the cyclic behavior of some of the solar proxies are connected with geomagnetic activities and have implications for space weather. Hence, in recent years efforts have been made using machine-learning methods to forecast these solar proxies with a view to improve our understanding of “space weather”. Developing a comparable method for forecasting $p$-mode frequency shifts is therefore of interest for two reasons. Firstly, it will facilitate future investigations into its potential role in tracing energy drivers from the Sun’s interior to the geospace response by improving models of solar interior dynamics to coronal and heliospheric plasma conditions. In other words, it will help establish a more robust and quantitative link between the Sun’s interior and its exterior. Secondly, it may provide us with an independent indicator or an early indicator of ascending and descending phase of solar activity which might be useful for space weather forecasting.

In this article, we develop and apply the standard time-series analysis and machine-learning based methods to characterise $p$-mode frequency shifts for the remaining solar cycle 25.
\end{abstract}
\end{frontmatter}
\section{Introduction}
     \label{S-Introduction} 

The 11 year activity cycle of the Sun \citep{Schwabe1843} is characterized by many proxies such as sunspot numbers, 10.7~cm radio flux, flares, geomagnetic storms etc.  Ever since the sunspot-records began, the cycles have been numbered starting from solar cycle 1 for the years 1755-1766. The current solar activity cycle is solar cycle 25. These activity cycles are studied with great interest for understanding the Sun and their influence on space weather \citep[e.g.][]{Mursula2007, irene2014, Hajra2021, Baker2000}.  

The oscillation frequencies observed on the Sun’s surface also rise and fall in step with the Sun’s activity cycle. These oscillations, commonly referred to as p‑modes, are acoustic resonances within the Sun, driven by turbulent convection in its convection zone. Their main restoring force is pressure gradient. 

These acoustic oscillations were first detected by \citet{Leighton1962}, \citep[see also,][]{Ulrich1970}. However, the variation in frequencies due to changes in magnetic flux were reported much later in 1985 by \citet{Woodard1985}. It is now well established that the $p$-mode  oscillations  of low-degree modes have a slightly higher frequencies upto $\sim$ 3.7 mHz, during the increased solar magnetic activity compared to when the magnetic activity is low \citep[see for example,][]{Libbrecht1990, Elsworth1990}. Above $\sim$ 3.7 mHz, a downturn is observed \citep{Salabert2004} and the frequency shifts start to decrease with even a reversal in sign above the acoustic cut-off \citep{Rhodes2002}. The increase in the frequency shifts below the $\sim$ 3.7 mHz suggest changes near the photosphere with solar activity. The decrease above 5.0 mHz are likely due to changes in the chromosphere.  Theoretical models have been developed with a view to understand the physical mechanism for such rise and fall in the frequencies. For example, \citet{Goldreich1991} have suggested changes in the mean square magnetic field strength in the photospheric region. However, as pointed out by these authors such mean field strength perturbations necessary to explain the observed frequency shift would require magnetic flux tubes to continue to strengthen to larger depths than expected. \citet{jain1993, jain1996} attributed the observed frequency shifts as a consequence of both magnetic and thermal changes, the ascending phase due to an increase in the mean photospheric magnetic field and the descending phase from a combination of an increase in mean chromospheric magnetic field strength and an increase in chromospheric temperature. Such a combination is able to produce results similar to what is seen in the observations but the required temperature changes in their model seem quite large.  Although the physical cause of the perturbation has not yet been firmly established, $p$-mode frequency shifts are believed to be caused by the changes in the plasma properties through which the waves propagate or due to direct effect of magnetic fields on the waves through Lorentz force or both \citep{Goldreich1991, jain1993}.  In addition to the frequencies, the mode line widths and amplitudes have been found to vary with magnetic activity cycle \citep[][and references therein]{kiefer2018}.

It is now well established that magnetic activity observed at the solar surface is intimately coupled to the physical processes and dynamical conditions operating within the solar interior. However, there is no full consensus on how exactly magnetic fields are generated in the interior, transported to the surface and then some magnetic flux  gathering into organised and compact sunspots. Even the location and precise nature of the dynamo is still under debate \citep{Charbonneau20}. Some theories suggest that the dynamo site is located at the base  of the convection zone, generally referred to as tachocline. The position and the physical conditions of the base of the convection zone have been shown to change with the solar cycle \citep{Basu2021}. As mentioned earlier, it is also argued that there is another site of dynamo, in addition to tachocline which is operating near the NSSL \citep{Benevolenskaya1998, Jain2022}. 
The inference of two distinct dynamo sites is further motivated by the presence of shorter‑period variations—on the order of 2–3 years—in helioseismic observations \citep{Fletcher2010, Mehta2022, Jain2023} and other solar activity indices \citep[for a review, see][]{Broomhall2014}, superimposed on the dominant 11‑year cycle. These shorter‑term modulations, known as quasi‑biennial oscillations (QBOs), have gained increased attention because such periodicities may provide a significant diagnostic for evaluating solar dynamo models.  

Since both, frequencies of $p$-mode oscillations and the strong magnetic flux regions seen on the solar surface have their origin in the interior and show cyclic and in-phase variation with time, it raises an important question: do strong magnetic‑flux regions contribute more significantly to the observed $p$-mode frequency shifts than other sources of perturbation?
It is therefore, interesting to examine other perturbations that connect $p$-mode properties and magnetic activity cycle.  Frequencies of solar $p$-modes have been used in measuring changes in the large-scale flows. The meridional flows that transport near-surface material towards the poles are an essential component of some solar-dynamo theories and may play a role in filtering dynamo models. It has been shown from helioseismic measures that meridional-flow speed changes with the Sun's magnetic activity cycle.  Torsional oscillations which are helioseismic measures of near-surface zonal flows \citep{Howard1980, Howe2018} also vary with solar cycle. \citet{Sasha2019} suggest that variation in zonal flows may carry the signature of a dynamo wave generated at the base of the convection zone  (see also, \citet{Mandal2024}).
Thus, changes in $p$-mode frequencies with solar cycle may hold keys for constraining solar dynamo models. 

The variation of solar $p$-mode frequencies are revealing information about various aspects of flows and magnetism in the Sun and are therefore, important to study. In this article, we forecast variation of $p$-mode frequency shifts for solar cycle 25 by using standard time-series analysis techniques such as wavelet and a combination of locally estimated scatterplot smoothing (LOESS), Fast Fourier Transform (FFT). Deep learning method with neural basis expansion analysis (NBEATS) is also used separately. In the wavelet transform method, we use pywt.waverec  (function in python library PyWavelets) to extract low frequency trend and high frequency noise. After smoothing the trend, we use continuous wavelet transform pywt.cwt  (function in python library PyWavelets) to search for repeating patterns both at short and long term cycles. For an independent comparison, we also examine the predictions by using a combination of LOESS and FFT methods, the figures for which are denoted by LOESS$\_$FFT. The 
LOESS$\_$FFT method involves combining LOESS for smoothing and FFT for frequency analysis. The prediction is then the average of the two. For nonlinear pattern fitting in both cases, we use the Light Gradient-Boosting Machine (LGBM). Deep learning methods are also known for powerful learning capabilities trained on long time series to fit complex nonlinear relations. \citet{Su2023} recently used this method with the neural basis expansion analysis (NBEATS) for solar cycle prediction. We also employ this method for the longer time series of sunspot numbers and Radio flux. These are applied to three different datasets: (i) $p$ mode frequency variations measured in solar cycles 23, 24 and 25,  i.e. 7 May 1995 to 31 July 2025 (ii) prediction of 10.7~cm radio flux using the flux dataset from 9 January 1954 to 31 July 2025 and then forecasting the $p$ mode frequency shifts using its best-fit relationship with the flux
(iii) prediction of sunspot numbers using the sunspot number (SSN) data from 9 January 1954 to 31 July 2025 and then forecasting the $p$ mode frequency shifts using its best-fit relationship with the SSN. In Section~2, we describe the datasets. In Section~3, we briefly mention the data analysis techniques that are used here. The various results obtained from the three datasets are then shown in Section~4 with Discussion and Conclusions in Section~5.

\section{Data}
     \label{S-Data} 
We use {\it p}-mode frequency data from Global Oscillation Network Group (GONG). GONG is a ground based network that measures Doppler velocities of the solar surface from six different stations located around the world \citep{Haarvey1996}. GONG has been providing coherent data for mode frequencies from 1995 to now. The GONG data analysed here covers the period from 7 May 1995 to 31 July 2025; this period covers two complete solar cycles 23 and 24 and part of the solar cycle 25. 

GONG routinely provides 108‑day mode frequency tables with 36‑day overlaps. To investigate short‑term variations, we instead use frequencies computed from nine‑day time series corresponding to the Legendre's polynomial expansion\footnote{https://gong2.nso.edu/ftp/TSERIES/v1z/}. Details of the frequency‑extraction procedure are described in \citet{Tripathy2007}. Because the measured mode frequencies depend on the mode mass, we follow the method of \citet{Woodard1991} and define the mean frequency shift ($\delta\nu$) as the mode‑inertia–weighted sum of the individual frequency measurements,
\begin{equation}
\delta\nu(t)\,=\,{\sum_{n,\ell}\frac{Q_{nl}}{\sigma_{n,\ell}^2}} \delta\nu_{n,\ell}(t)
/\sum_{n,\ell}\frac{Q_{n,l}}{\sigma_{n,\ell}^2} ,
\end{equation}
\noindent
where $\sigma_{n,l}$ is the uncertainty in the frequency measurement for a given mode defined by the radial order $n$ and spherical harmonic degree $\ell$, $\delta_{n,\ell}$(t) is the change in a given mode, and $Q_{n,\ell}$ is the mode inertia ratio defined by \citet{jcd91}. The reference frequency is chosen to correspond to the minimum value in the time series. The analysis uses common modes with frequencies in the range $1500 \le \nu \le 5500\; \mu$Hz and degrees spanning $0 \le \ell \le 100$. 

Since the discovery of the relationship between solar radio emissions and sunspot numbers in 1947 by \citet{covington47}, 10.7~cm radio flux (hereafter referred to as F10) has been used as a well known indicator of solar activity. 
F10 corresponds to a wavelength of 10.7~cm and gives a measure of the total daily emissions of radio waves within frequencies $2700 \le \nu \le 2900$ MHz, observed from the solar disk\footnote{https://www.spaceweather.gc.ca/forecast-prevision/solar-solaire/solarflux/sx-5-en.php} \citep{Tapping2013} and is measured in solar flux units (sfu), where 1 sfu =$10^{-22}$Wm$^{-2}$Hz$^{-1}$. It mainly represents the  contributions from sunspots and radio plages in the upper
chromosphere in addition to the quiet-Sun background
emission \citep{Covington1969}.

The location and the number of sunspots observed on the solar surface vary as solar cycle progresses. These are recorded daily in both hemispheres from the visible part of the solar disk\footnote{www.sidc.be/silso/datafiles}  \citep{SILSO_Sunspot_Number}.

\begin{figure}
\includegraphics[width=12cm]{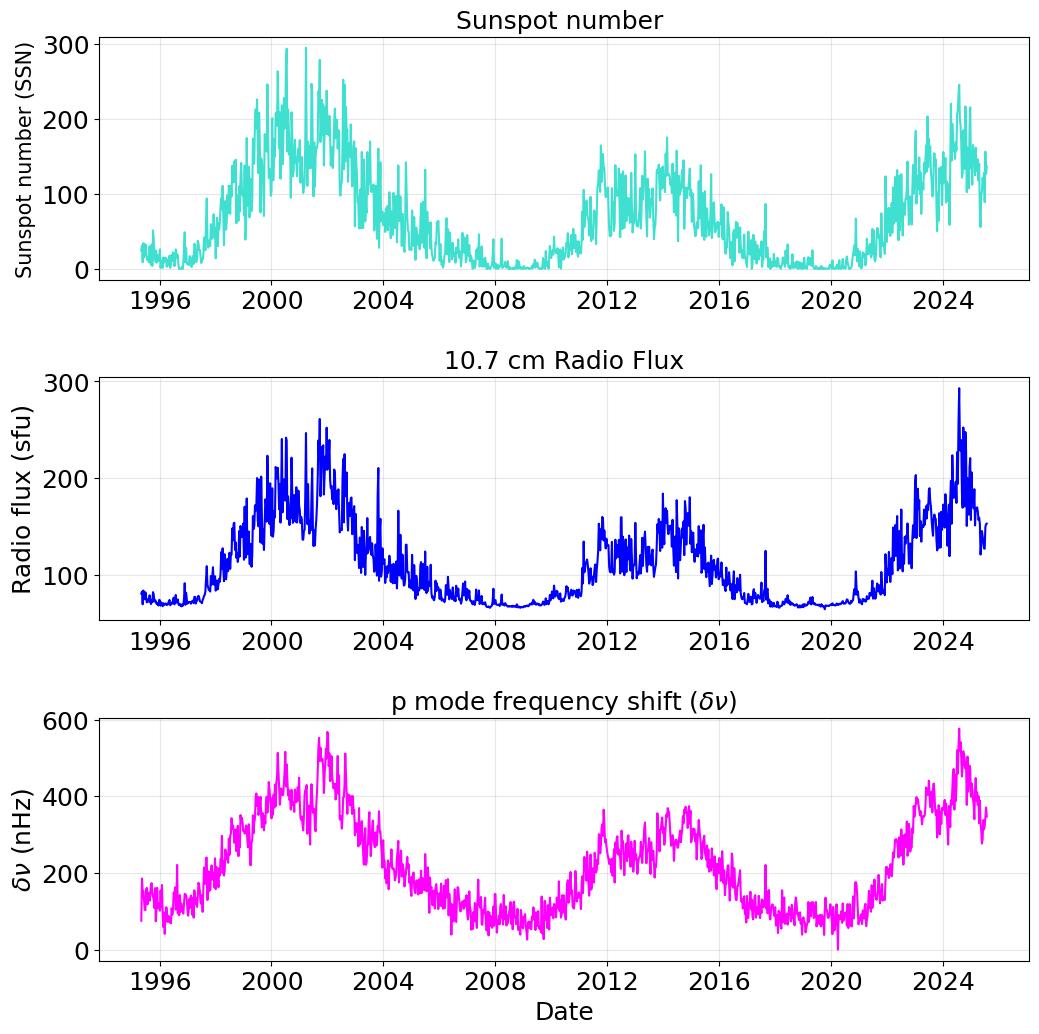}\hfill
\caption{Temporal variation of Sunspot numbers, 10.7~cm radio flux and $p$-mode frequency shifts, all averaged over the same 9 day period.}
\label{fig:Fig1}
\end{figure}

Figure \ref{fig:Fig1} displays the Sunspot numbers (top), 10.7~cm radio flux (middle) and $p$-mode frequency shift,  $\delta\nu$ (bottom) as a function of time starting from May 1995. All three dataset shown are averaged over the same 9 day intervals. It is evident that all these quantities exhibit similar temporal variation. 

\section{Data Analysis Techniques}
     \label{S-Data Analysis Techniques} 
\begin{figure}
\includegraphics[width=6.0cm]{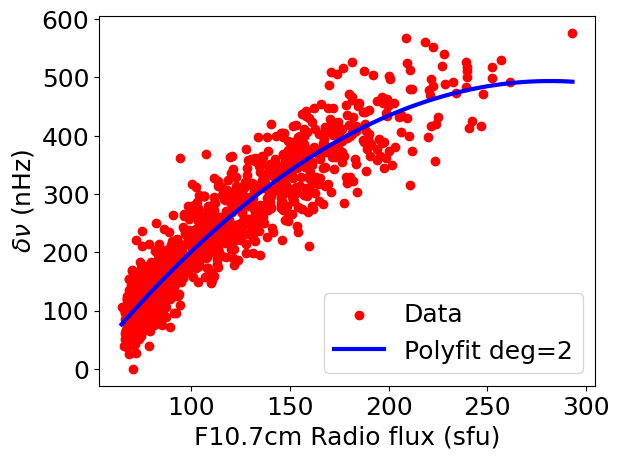}
\includegraphics[width=6.1cm]{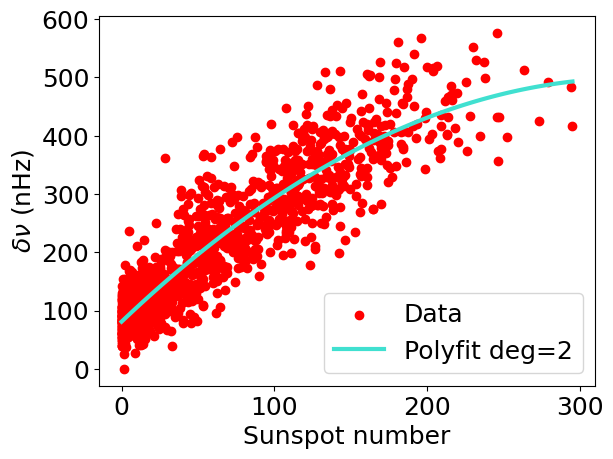}\hfill
\caption{Relation between the $p$-mode frequency shifts and F10 (left) and sunspot numbers (right).}
\label{fig:correlation}
\end{figure}

 The $p$-mode frequency shifts have long been known to correlate with 10.7~cm radio flux \citep{Woodard1985} and the sunspot numbers \citep[see][]{Rhodes91}. The data we use for the radio flux and sunspot numbers recorded are averaged over a 9-day time period to match up with the GONG time series for the $p$-mode frequencies.  Note that the sensitivity of the $p$ mode frequencies to the change in the magnetic activity and other activity proxies have also been explored and found to be similar with other instruments \citep[see for example][]{Salabert2004, Cristina2006, Chaplin2007, Jain2009, Larson2018} making them quite robust.  Figure \ref{fig:correlation} displays frequency shifts and the 10.7~cm radio flux (left panel) and sunspot numbers (right panel) with red filled circles. While most of the studies involving $p$-mode frequency shifts and solar activity indices report high linear correlation, deviations from a simple linear relation has also been reported  indicating a saturation effect at high solar activity \citep{Chaplin2007, Cristina2006, Cristina2019}. Since Figure~2 also suggests a saturation effect during periods of maximum  activity, we fit a  polynomial of degree 2 to each dataset. The resulting fitting  is displayed as a blue solid line (left panel) for F10 and a turquoise solid line (right panel) for the sunspot numbers. It is clear that the quadratic fit reproduces the higher‑magnitude data more accurately than the linear fit (the deviation is small); therefore, we adopt the quadratic form for subsequent analysis.
 
 The relationship is calculated only for datasets from May 1995 onwards since the $p$-mode frequencies from GONG are only available for this period. We will assume that this relationship is also valid before 1995. On that basis, even though the F10 data is available for a longer period starting from  February 1947, we use  datasets of F10 and SSN, from 9 January 1954 (since the minimum of the cycle starts around this period) to 31 July 2025 to forecast the $p$-mode frequency shift in the current solar cycle 25.  

The three parameters also have different kinds of errors as they are measured by different instruments; only data points from $\sim$ 2.5 cycles can (at this stage) be used for the best fit. There is no way to quantify the scatter in a meaningful way, so no error bars are shown in the figures and the results should be interpreted with caution. %interpretation requires caution. 
The fits are expected to improve as more data will become available in the future. Since we are interested in qualitative behaviour, we will proceed with the quadratic fit that we have obtained and use the following  techniques for forecasting.
\subsection{Machine Learning-based frameworks}
\label{S-MLFramework}
In order to predict the timing of the next minimum in the $p$‑mode frequencies, two methodologies have been applied to the time series processing (i) a standalone Wavelet-LGBM model (ii) an ensemble consisting of LOESS-LGBM and FFT-LGBM models. Both methodologies consists of signal decomposition, feature engineering techniques and model training and evaluation. 
\subsubsection{Signal Decomposition}
\label{S-Signal Decomposition}
The 9-day averaged time series of 10.7~cm radio flux and the sunspot numbers, each has a periodicity of approximately 11 years with dissimilar fluctuations.  As is customary, we decompose the signal into trend and residual components so that models can learn the temporal and seasonal dependencies more effectively. We applied three decomposition techniques to each of the time series. A brief description of each method in the context of our data is given below.
\\
\\
\textbf{\textit{Wavelet-LGBM:}} Wavelet transformation decomposes signals into multiple frequency components using localised wavelets, providing high resolution time and frequency information \citep[see][]{Shao2017}. In wavelet-LGBM model, as the name suggests a wavelet technique has been used for decomposition. The trend has been obtained using the Discrete Wavelet Transform (DWT) using `db4' \citep{Wasilewski2012}. The residual component is obtained by subtracting the obtained trend from the original timeseries.
\\
\vspace{0.1pt}

\noindent
\textbf{\textit{LOESS\_LGBM:}} LOESS (locally estimated scatterplot smoothing) is a non-parametric technique that fits low-order polynomials to extract smooth trends in both stationary and non-stationary signals. It has been used widely in utility forecasting \citep{Zhang21}, traffic prediction \citep[see for example][]{Chen2019, Zhao2022,  Yuan2025}, economic analysis \citep{Kehinde2020} etc. In LOESS\_LGBM model, we decomposed the signal into trend and residuals by using the optimally selected smoothing fraction based on the minimum error. For each smoothing fraction, the root mean squared error (RMSE) was calculated, and the fraction yielding the minimum error was selected as the final model parameter. The residual component is then obtained by subtracting the best fit trend from the original timeseries.
\\
\vspace{0.1pt}

\noindent
\textbf {\textit{FFT\_LGBM:}}  In this model, Fast Fourier Transform (FFT) is used as low pass filter to obtain the trend of the timeseries. It filters out spectral components exceeding 0.01 cycles/9 day and zeroing them, thus removing noise or very small fluctuations. Such a cut-off frequency enables us to retain variations longer than 2.5 years in this data set.  With these filtered spectral coefficients, Inverse FFT is used to reconstruct the trend component. The residual component is obtained in the same manner as performed for wavelet or LOESS-based decomposition.

\subsubsection{Feature Engineering}
The extracted components are further processed through feature engineering to incorporate temporal lags and seasonality, providing models with more context for better predictions.
\\
{\it Temporal Lags}: Lag features are generated for both the trend and residual components such that at each time t, each component was augmented with n previous observations ($t-1, t-2..,t-n$).
\\
\\
{\it Seasonality}: In Wavelet and Loess-based models, seasonality is accounted for by incorporating sinusoidal terms having a period determined by identifying the top spectral peaks in the wavelet power spectrum, which are then averaged to provide a mean period. In the FFT-based model, Power Spectral Density (PSD) was obtained using FFT to get the dominant frequencies. These are then converted into time period using $1/f$ and subsequently averaged to get the mean period for the seasonal term.
\\
\\
{\it Uncertainty Quantification}: Due to inherent uncertainties in the time series of $\delta\nu$, SSN or F10, prediction intervals are also created using LGBM-based quantile regression for all three models. Specifically, the 5th and 95th percentiles of the data are used to predict the lower and upper bounds of the forecast, respectively (creating a 90\% prediction interval).
\\
\\
{\it Model Training and Validation}: For each of the methods described in subsubsection~\ref{S-Signal Decomposition}, separate LGBM models are trained; one for the trend component, and another for the residual component.  For Wavelet-LGBM methodology, the final forecasted values are obtained by adding the predicted trend and the predicted residual value.
For ensemble methodology of LOESS\_LGBM and FFT\_LGBM, the forecasted values are obtained for each model by adding their predicted trend and residual value. Then, the final forecast ($Y_{ensemble}$) from LOESS ($Y_{LOESS}$) and FFT ($Y_{FFT}$) based models are averaged by averaging the predicted
\begin{eqnarray*}
Y_{ensemble}=\frac{Y_{LOESS} + Y_{FFT}}{2}.
\end{eqnarray*}
To preserve the temporal dependencies, k-fold validation  which involves shuffling, was avoided \citep{Su2023}. Instead hold-out validation strategy was employed using an 85:15 split ratio for the train and test data, ensuring the training dataset precede test dataset. Models’ performances were evaluated on test data using RMSE. Upon satisfactory results, models are retrained on the entire timeseries data to get future predictions.

\subsubsection{Deep Learning: N-BEATS}
We use deep learning N-BEATS method which is a multi-layered prediction method \citep{Oreshkin2019} and has two components: trend forecast component and residual forecast component. For trend forecast, stacks of ‘trend’ and
‘seasonality’ were specifically chosen as trend and periodic oscillations. For residual
forecast, since it is inherently noisy, stack of ‘identity’ is used to model the fluctuation. Forecasts from both the models are aggregated back to give complete forecast. Such a method has been recently implemented by \citet{Su2023} for the prediction of the variation of the sunspot numbers for solar cycle 25. We only use this method for $\delta\nu$ forecast using SSN and F10 correlations. We do not use this method to forecast $\delta\nu$ directly from $\delta\nu$ dataset as $\delta\nu$ dataset is not long enough to use deep learning method.
\section{Results}
\label{S-Results}
\subsection{Direct Prediction of $p$-mode frequency shift ($\delta\nu$)}
\begin{figure}
\includegraphics[width=12cm]{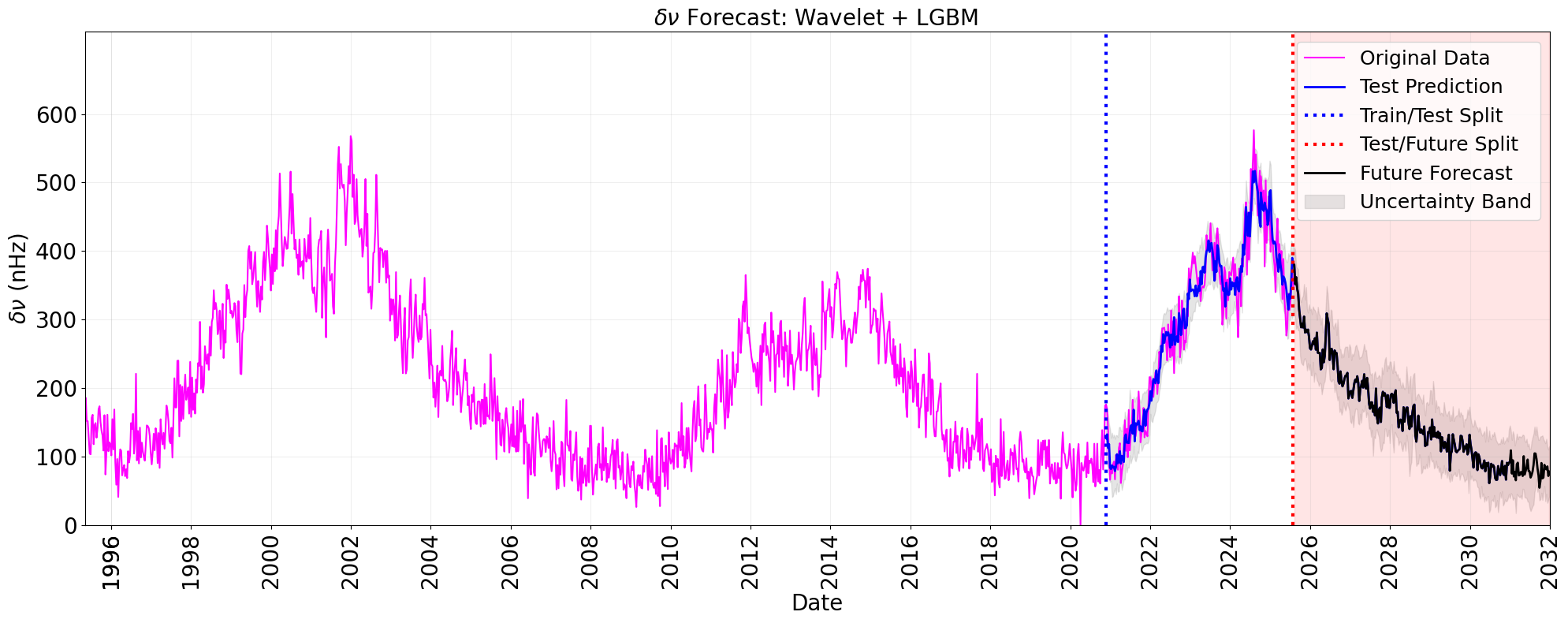}
\includegraphics[width=12cm]{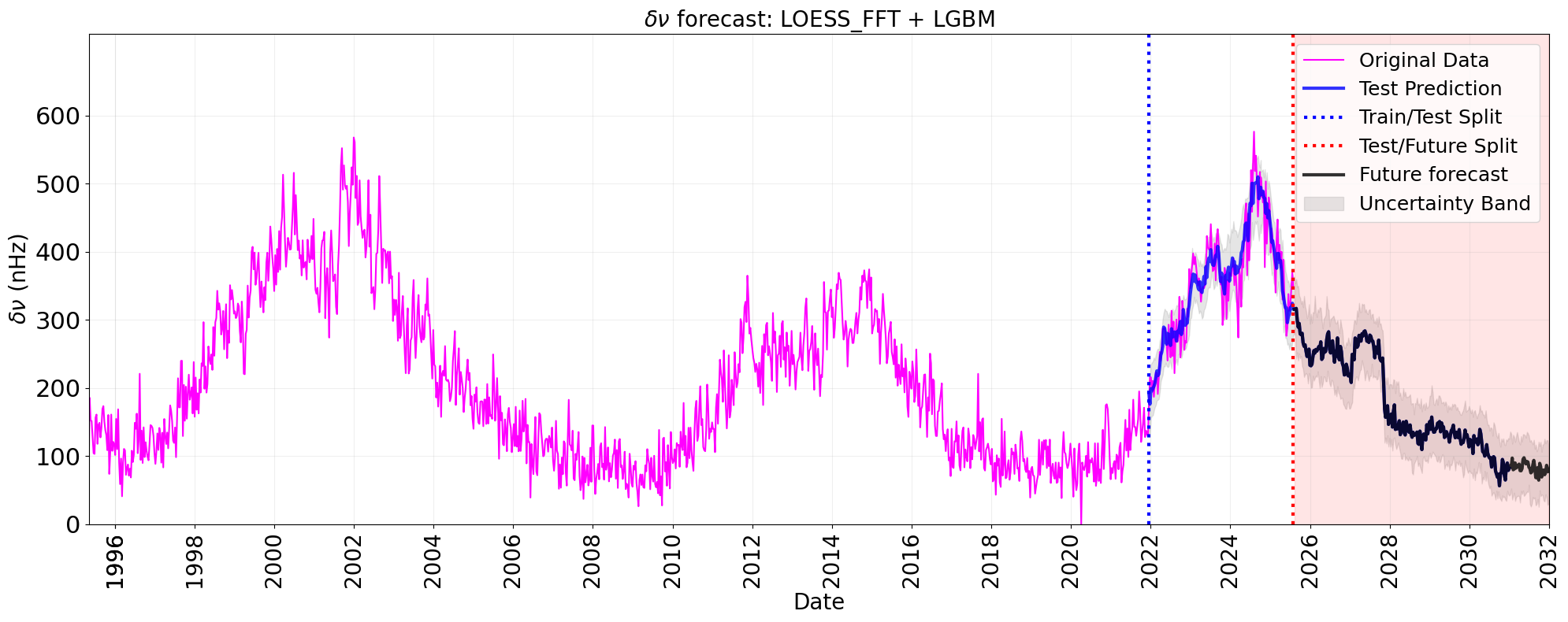}
\caption{Temporal variation of averaged $p$-mode frequency shift, $\delta\nu$. The pink shaded region displays the forecasted $\delta\nu$ with 90 percentile confidence level in grey. The top panel shows forecast using wavelet+LGBM, the bottom panel is obtained from using LOESS\_FFT+LGBM.  The forecasted $\delta\nu$ in all panels are obtained by using $\delta\nu$ data from 7 May 1995 - 31 July 2025.}
\label{fig:Direct}
\end{figure}

Figure \ref{fig:Direct} displays the temporal variation of frequency shift, $\delta\nu$. The pink shaded region displays the forecasted $\delta\nu$. These forecasted $\delta\nu$ are obtained from the original $\delta\nu$ dataset covering the period 7 May 1995 - 31 July 2025. The 95 percentile confidence level is shown in grey.  The top panel shows the full dataset, original observed dataset in magenta and forecasted $\delta\nu$ from Wavelet+LGBM technique in black. The bottom panel is same as the top except the method used is LOESS\_FFT+LGBM. The forecasted $\delta\nu$ suggests that the solar cycle 25 is already in the descending phase and the cycle minimum is expected in the year 2031. Since the original dataset comprises of only two complete solar cycles, solar cycles 23 and 24, and partial solar cycle 25, predictions beyond the next few years should be interpreted with caution. 

%there is a need for caution here, particularly regarding the interpretability of predictions beyond a few years.
\subsection{Prediction of $\delta\nu$ based on the relationship with 10.7 cm radio flux}
\begin{figure}
\includegraphics[width=12cm]{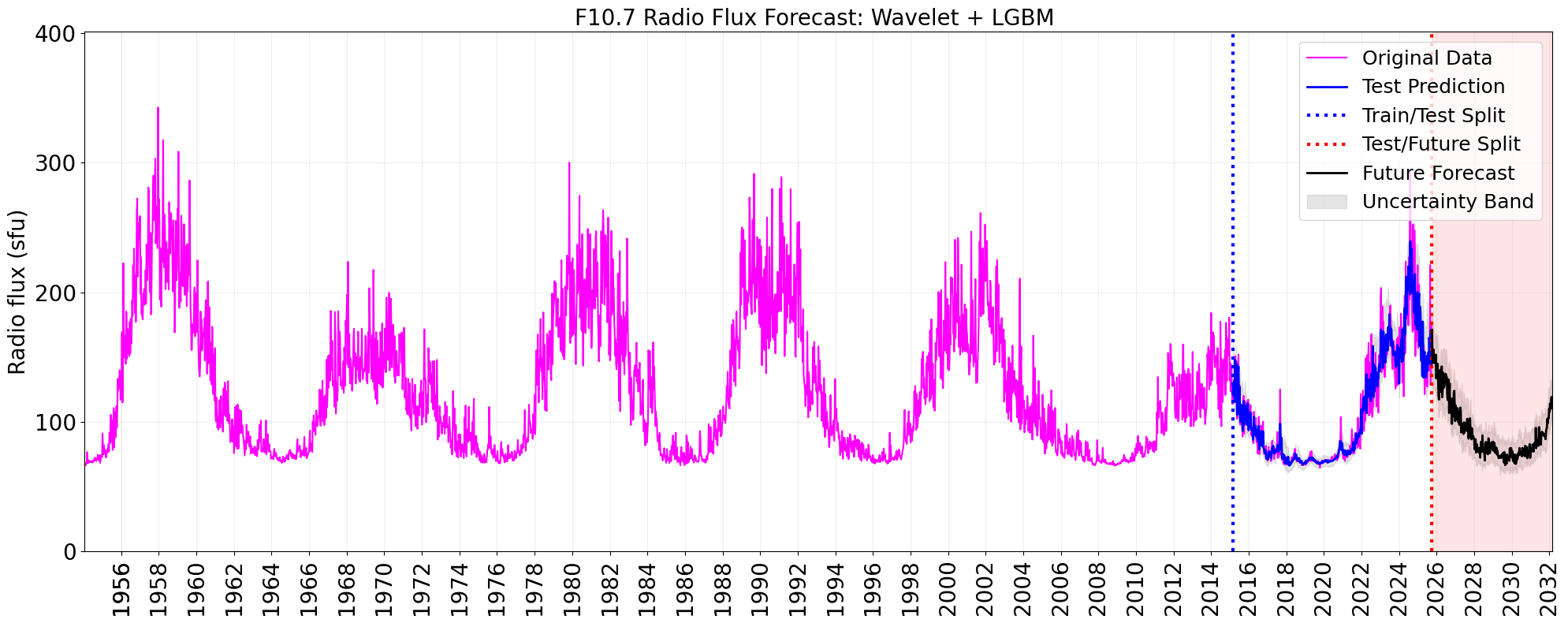}
\includegraphics[width=12cm]{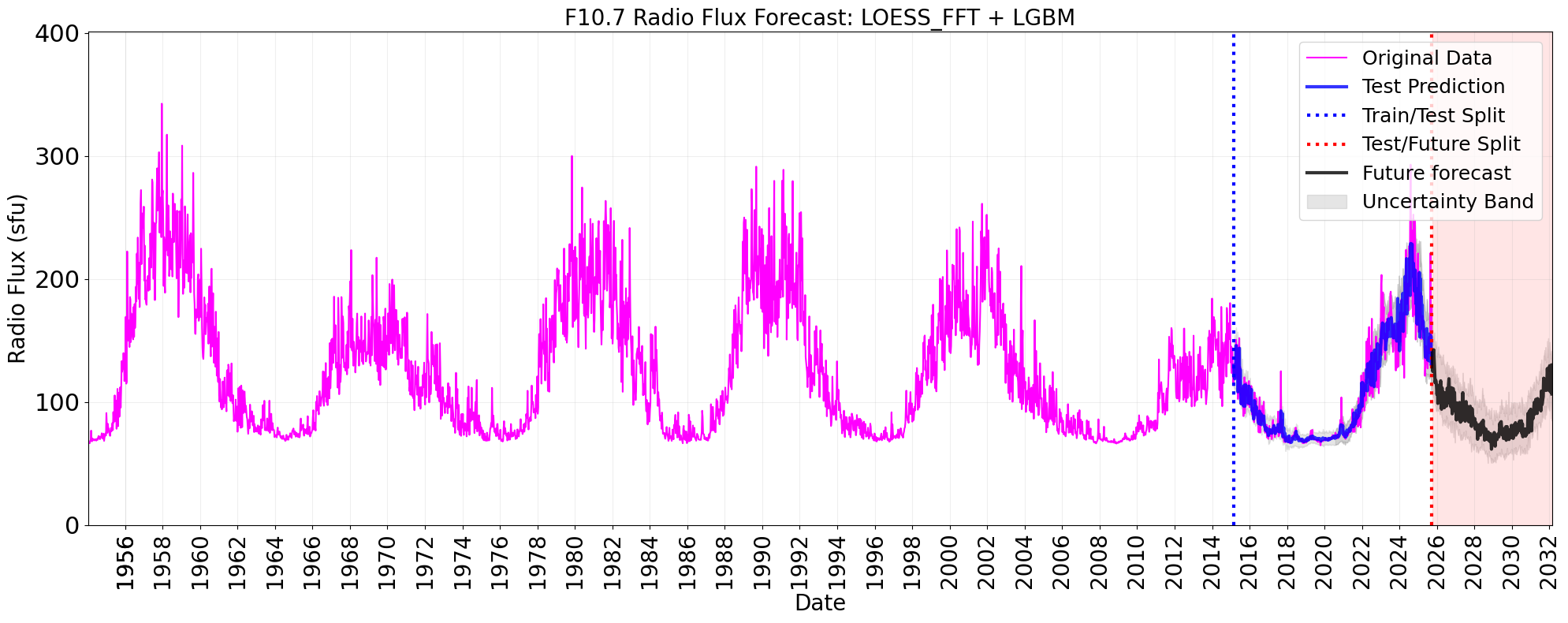}
\includegraphics[width=12cm]{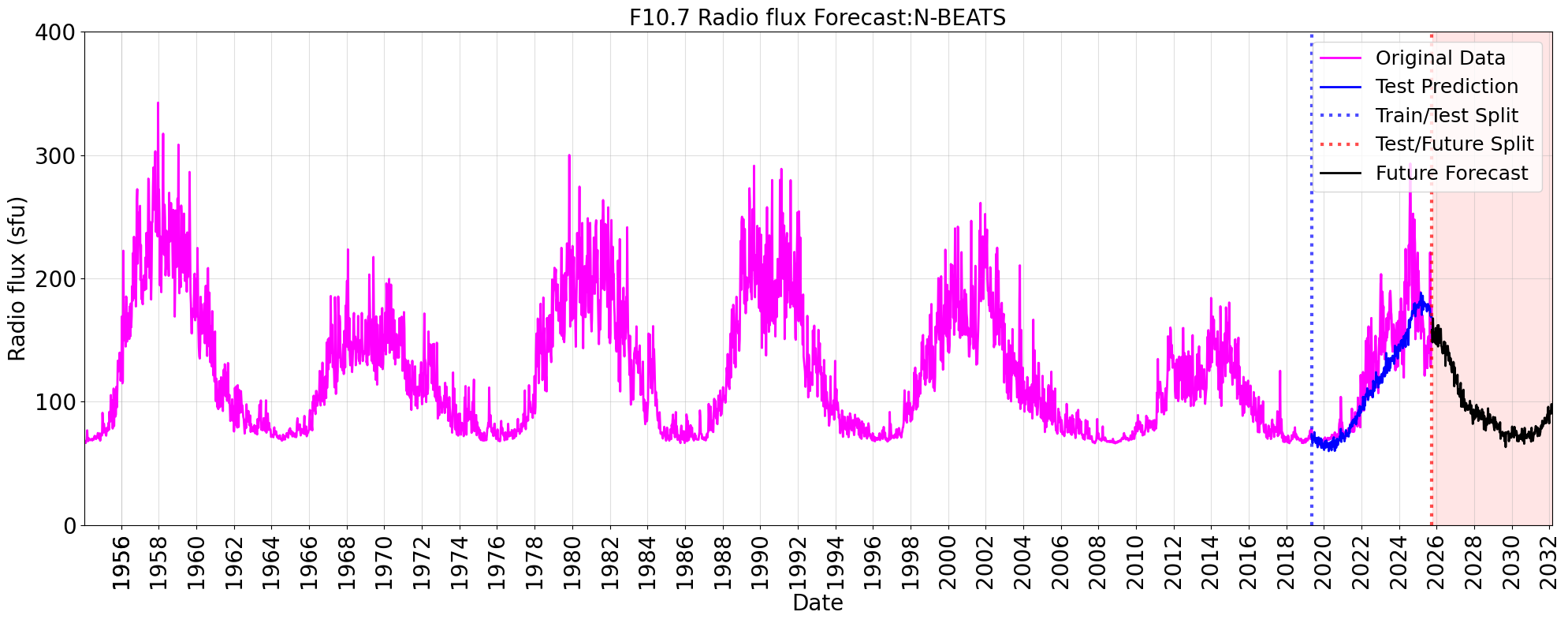} 
\caption{Forecast of 10.7~cm radio flux using Wavelet+LGBM technique (top), LOESS\_FFT+LGBM technique (middle) and NBEATS (bottom).}
\label{fig:wavelet+Loess+FFT}
\end{figure}
\begin{figure}
\includegraphics[width=12cm]{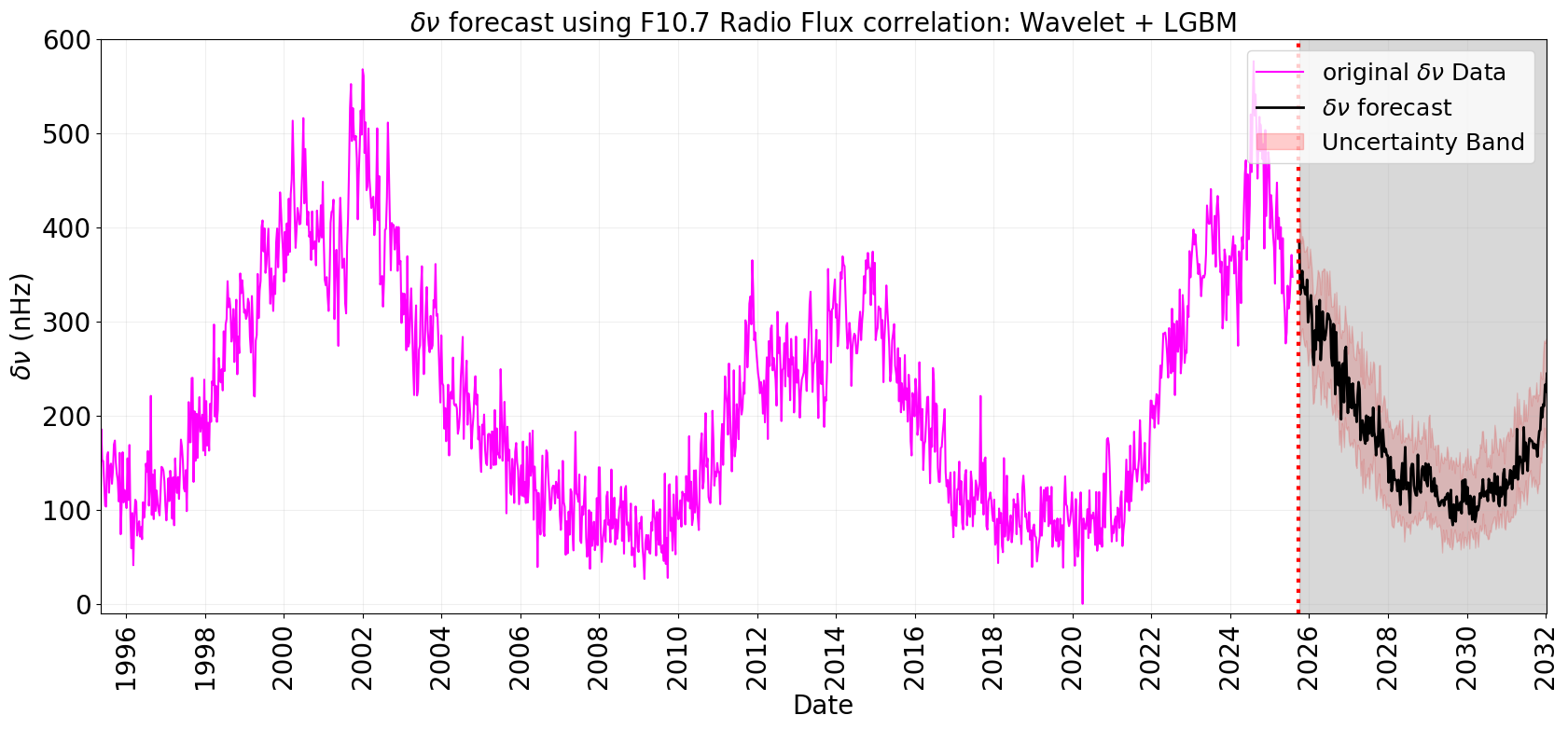}
\includegraphics[width=12cm]{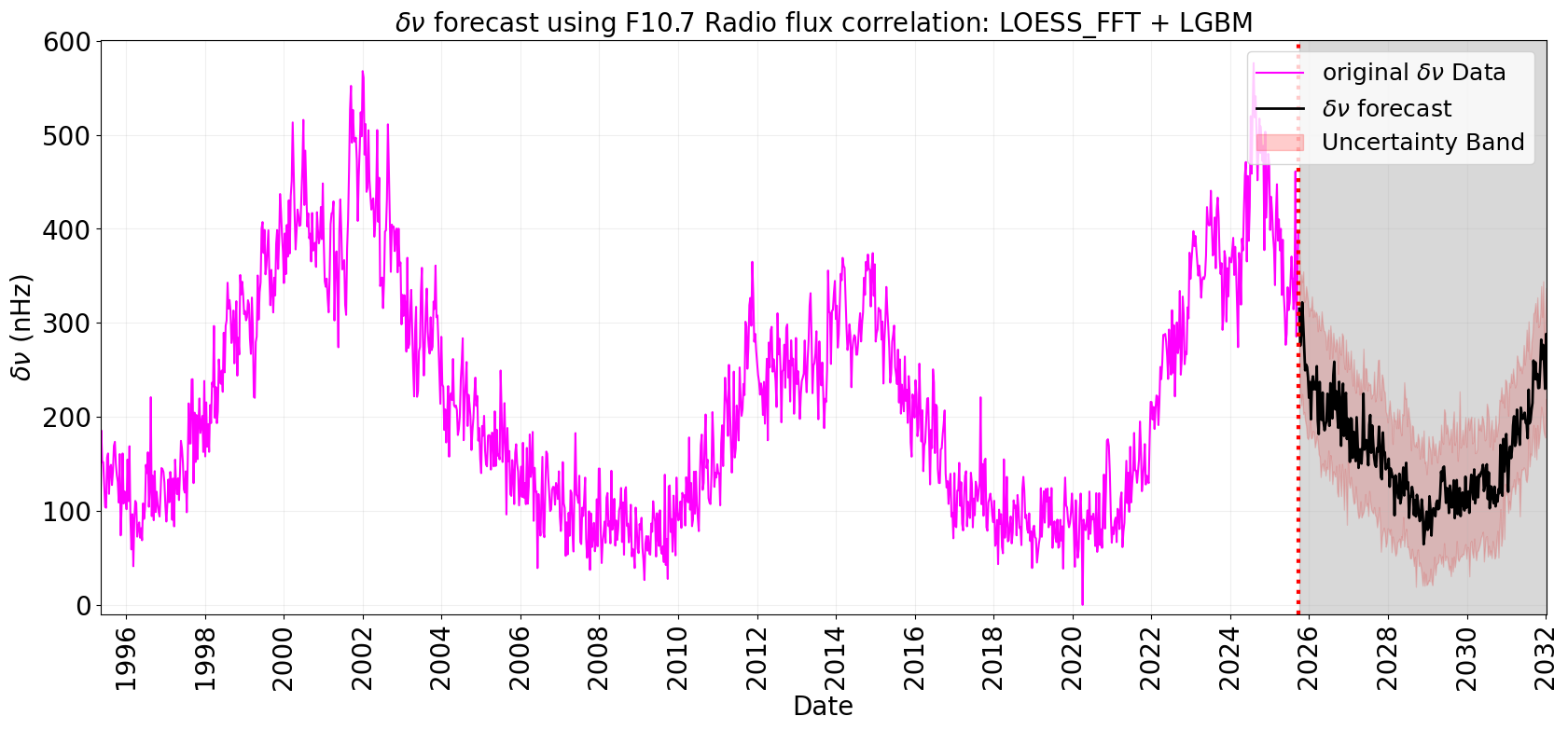}
\includegraphics[width=12cm]{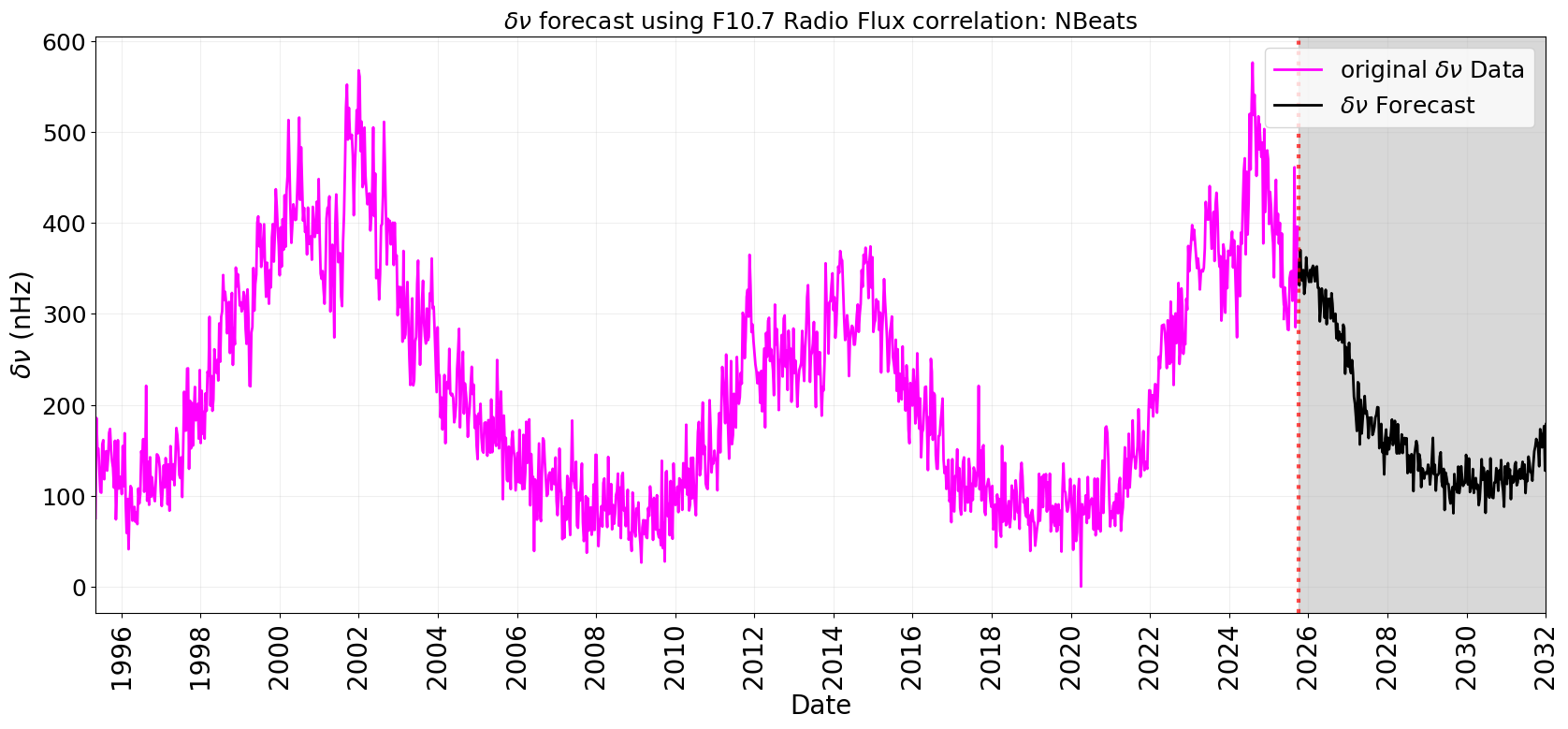}
\caption{Scaled-forecast of $p$ mode frequency shift using the relationship between F10 and $\delta\nu$. The forecast is obtained using Wavelet+LGBM technique (top), LOESS\_FFT+LGBM technique (middle) and NBEATS (bottom).}
\label{fig:dnu_RF}
\end{figure}
Here, we first find the prediction of the 10.7~cm radio flux and then compute the scaled-forecast of $\delta\nu$ from this prediction.
Recall that the radio flux data used in this study covers the period from January 09, 1954, to July 31, 2025. Note that each data point represents the average value over a consecutive 9-day interval. The forecast is obtained using various methods: wavelet+LGBM, LOESS\_FFT+LGBM and NBeats. These are shown in Figure \ref{fig:wavelet+Loess+FFT}. The vertical dashed blue line separates the training (magenta) and test (blue) dataset. The test dataset is in very good agreement with the original data in the top and middle panel. The NBeats method, shown in the bottom panel, underestimates the ascending part. However, the predicted F10 displayed here in black within pink shaded area is qualitatively similar in all three panels.  We, then make use of the polynomial fit obtained in Figure \ref{fig:correlation} along with these forecasts for F10 to compute the scaled-forecast for frequency shifts $\delta\nu$. The computed scaled-forecast is displayed in Figure \ref{fig:dnu_RF} in the grey shaded region with black solid line. The top, middle and bottom panels correspond to wavelet+LGBM, and LOESS\_FFT+LGBM and NBEATS methods respectively. The pink shaded region around the black curve indicates 90 percentile confidence interval. 
It is apparent from Figure \ref{fig:dnu_RF} that the fluctuations in the scaled-forecast is different for different methods. The wavelet+LGBM (top) shows smoother descent compared to LOESS\_FFT+LGBM (middle) which is steeper. The NBeats method (bottom) shows slightly different decline to the other two but all three methods suggest a minimum $\delta\nu$ around 2030.

\subsection{Prediction of $\delta\nu$ based on the relationship with Sunspot number}
%Number of sunspots recorded from January 1749–present are available in public domain through SILSO, World Data Center, Royal Observatory of Belgium -www.sidc.be/silso/datafiles.

In order to compare the scaled-forecast of $\delta\nu$ between the SSN and F10, we use SSN dataset from the same period i.e. from January 09, 1954 to July 31, 2025. We also average SSN within this period over the same consecutive 9-day interval as F10 and the $p$ mode frequencies. 
\begin{figure}
\includegraphics[width=12cm]{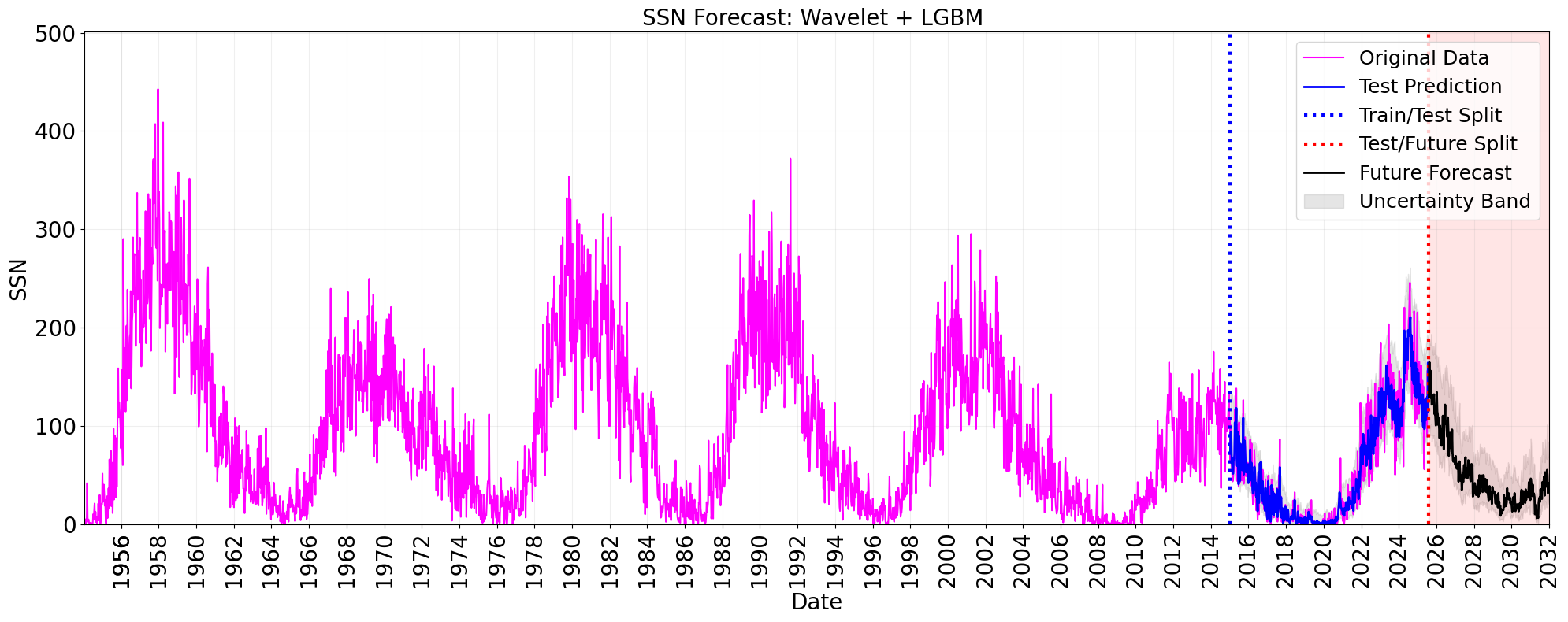}
\includegraphics[width=12cm]{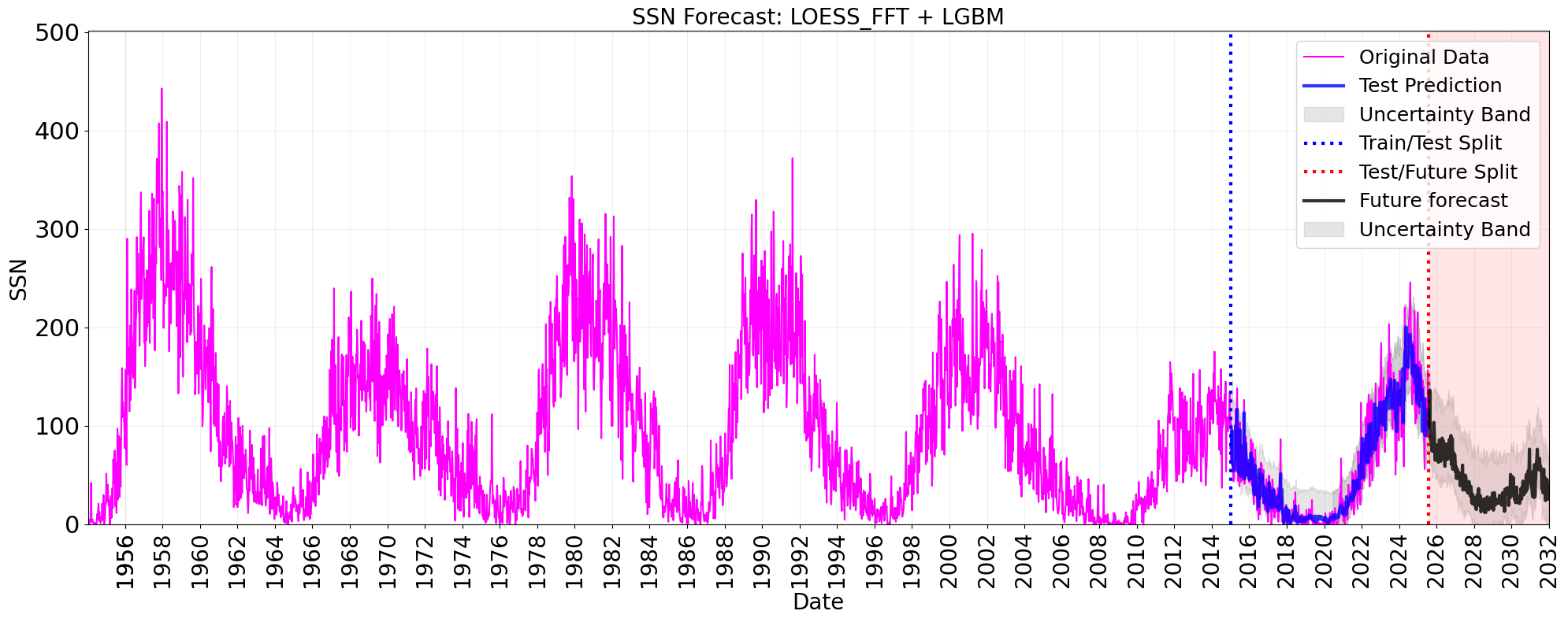}
\includegraphics[width=12cm]{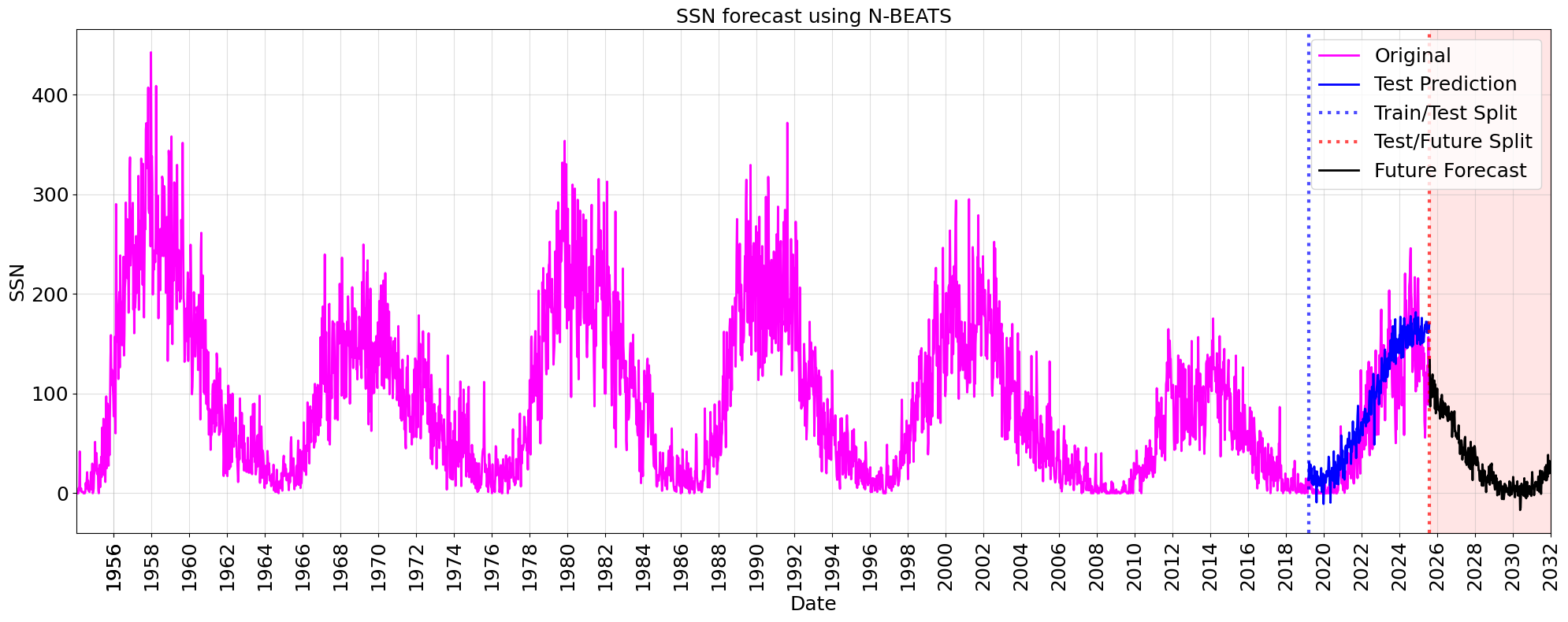}
\caption{Forecast of sunspot numbers using Wavelet+LGBM technique (top) and LOESS\_FFT+LGBM technique (middle) and Deep learning-NBEATS method (bottom).}
\label{fig:SSN_predict}
\end{figure}
Figure \ref{fig:SSN_predict} shows the forecasted SSN in SC25. Using the correlation of SSN with $p$-mode frequencies from Figure 2, we now compute the scaled-forecast of $\delta\nu$ and display it in Figure \ref{fig:dnu_SSN}. The three methods yield scaled-forecast which is different in detail. The minimum is between 2030-32 in the top and bottom panel but the middle panel has slightly larger fluctuations. Overall, the minimum is similar ($\sim$ 2030) to the one obtained using F10 shown in Figure \ref{fig:dnu_RF}.
\begin{figure}
\includegraphics[width=12cm]{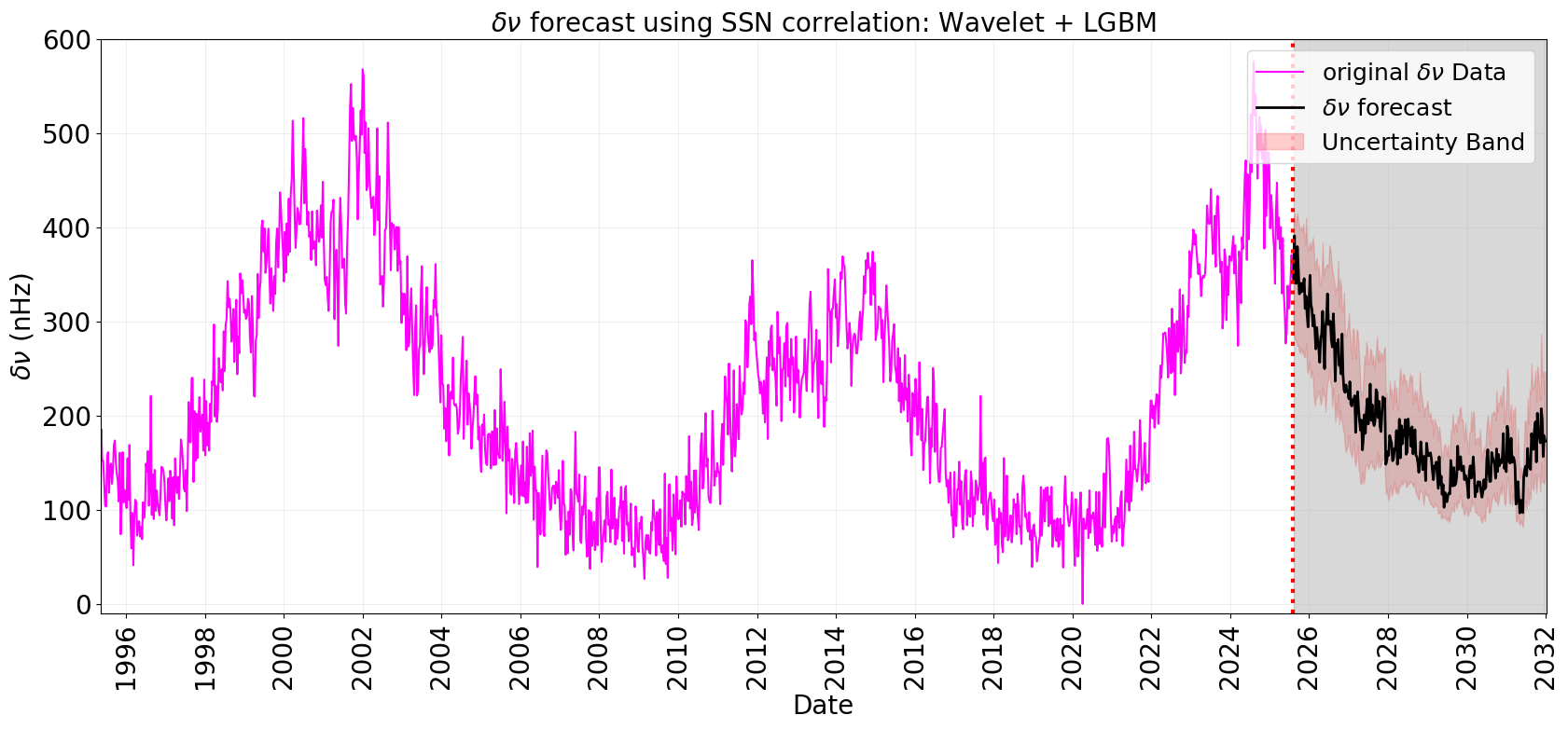}
\includegraphics[width=12cm]{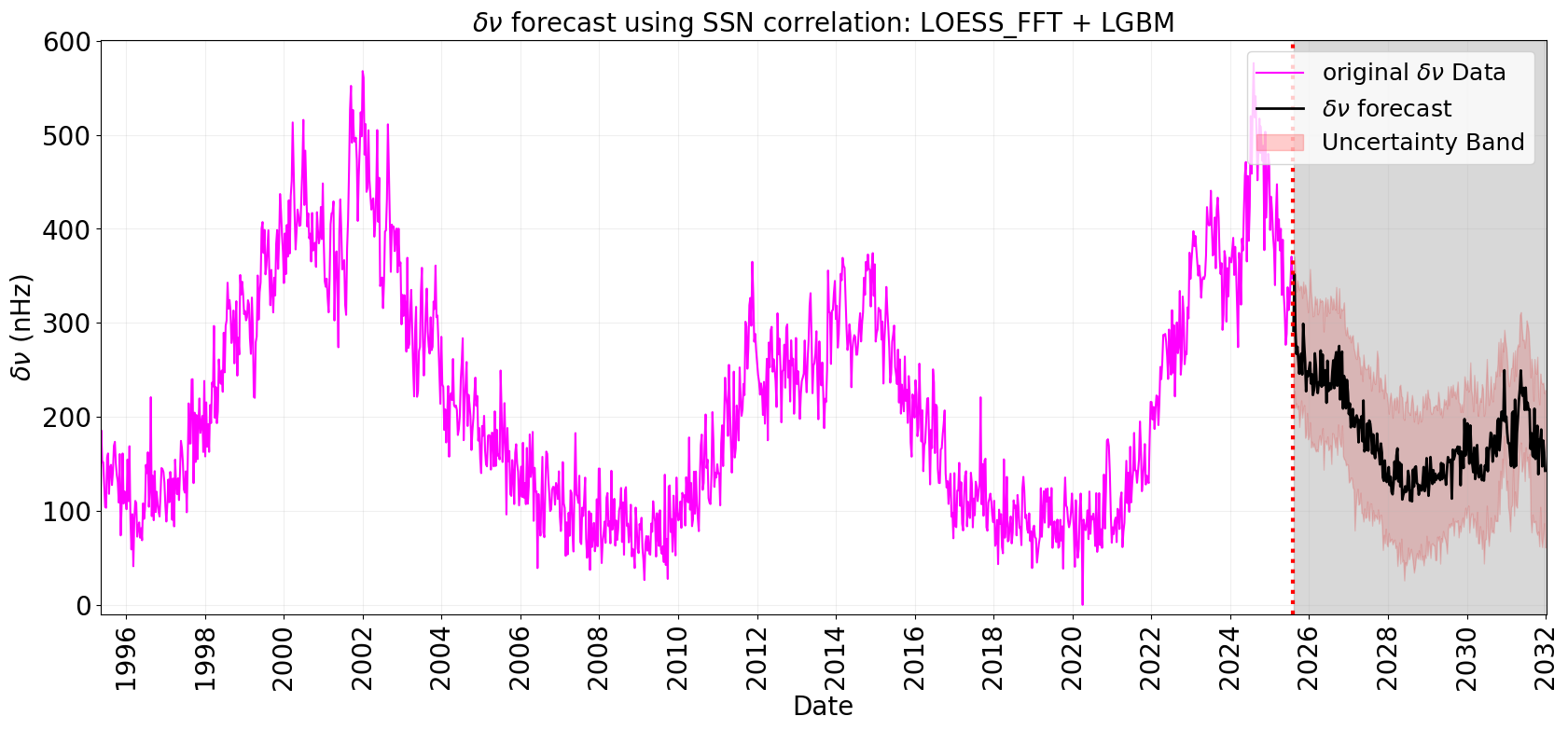}
\includegraphics[width=12cm]{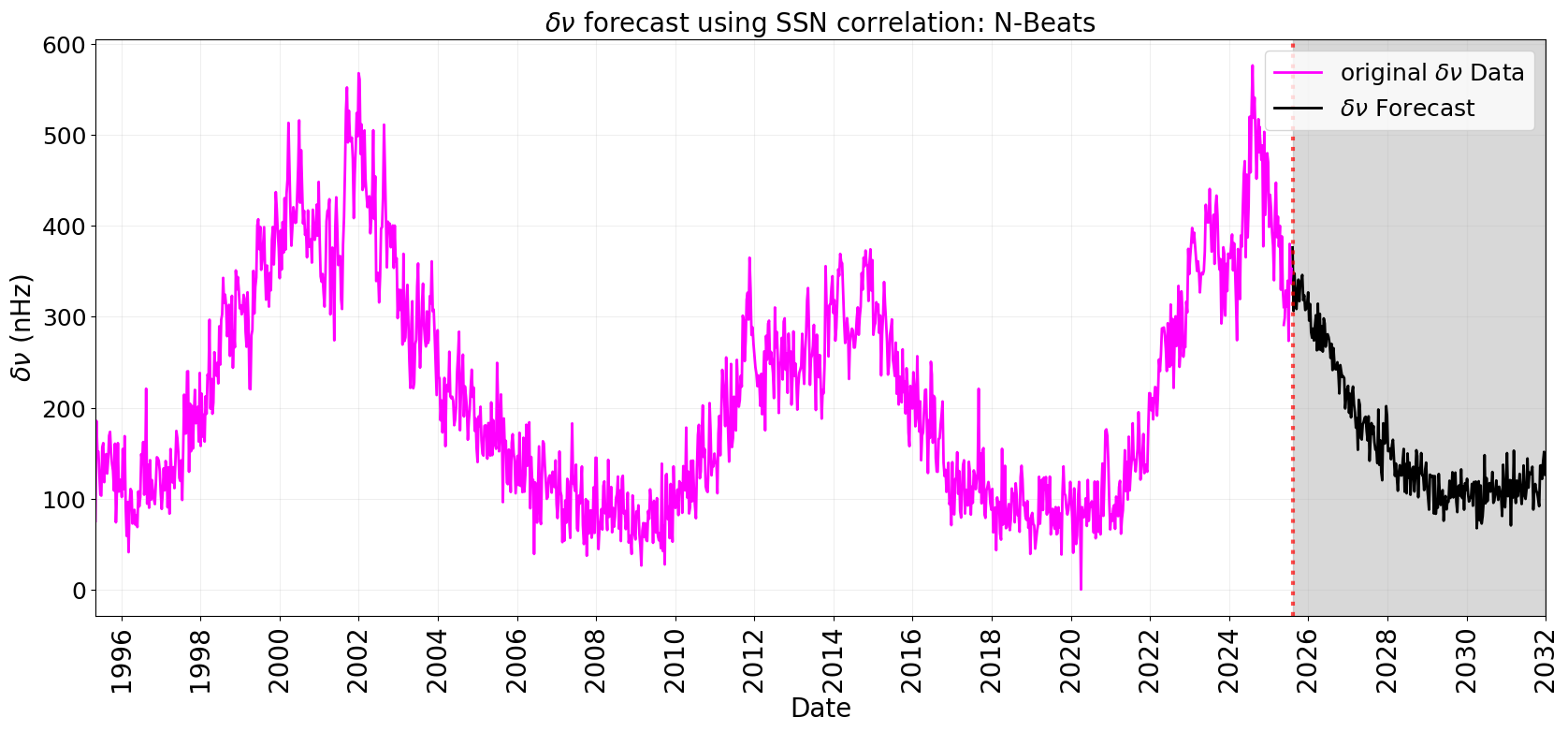}
\caption{Scaled-forecast of $p$ mode frequency shift using the relationship between SSN and $\delta\nu$. The forecast is obtained using Wavelet+LGBM technique (top), LOESS\_FFT+LGBM technique (middle) and NBEATS (bottom).}
\label{fig:dnu_SSN}
\end{figure}

\begin{table}[t]
\begin{center}
\begin{tabular}{ |c|c|c|c|c| } 
 \hline
 Sunspot cycle number & $\delta\nu$ minimum period  & $\delta\nu$ maximum period \\ 
 \hline
23 & March 1996 & January 2002 \\ 
 \hline
 24 & September 2009 & October 2014 \\ 
 \hline
  25 & February 2020 & August 2024 \\
  \hline
  26 & \textcolor{blue}{March 2030 (scaled forecast)} & \\
  &  \textcolor{blue}{January 2031 ($\delta\nu$ directly)} & \\
   \hline
 \end{tabular}
 \caption{The expected year of minimum of $p$ mode frequency shift in sunspot cycle 26, based on our analysis, is between 2030 and 2031 from direct $\delta\nu$ forecast and 2030 from scaled-forecast of F10 (see Figures \ref{fig:dnu_RF}) as displayed in blue.}
\end{center}
\end{table}

\section{Discussion and Conclusion}

Solar $p$ mode frequency shifts exhibit $\sim$11 year periodicity similar to many magnetic and related activity cycles. This is evident for solar cycle (SC) 23  and 24. In this paper we forecast the frequency shifts for the remaining part of solar cycle 25 by making use of the standard time-series analysis and machine-learning methods. We obtained forecasts in three different ways: (i) direct forecast using measured frequency shifts of solar cycles 23, 24 and 25 (partial), (ii) scaled-forecast using the best fit with 10.7~cm radio flux data and (iii) scaled-forecast using the best fit of frequency shifts with sunspot numbers. From figures \ref{fig:Direct}, \ref{fig:dnu_RF} and \ref{fig:dnu_SSN}, it is obvious that the descending phase of SC25 lasts for a longer period of time in (i) when compared to (ii) and (iii). This is likely due to fewer number of data points in (i). The time series used to directly forecast $p$-mode frequency shift, from 1995-2025, is too short for the next 11 year's prediction and therefore, not accurate enough to reliably point out the exact month and year of the minimum between SC25 and SC26. Based on Figure \ref{fig:Direct}, an approximate time period of 2031 can be seen here.  The scaled-forecast from F10 and SSN yields the minimum in the year 2030. \citet{Xu2024} likewise predict a comparable minimum period for SC26 for monthly sunspot numbers.  
 In Table 1, we mention the minimum period of $\delta\nu$ for cycles 23, 24, 25 from the original data of $\delta\nu$ as shown in magenta color in Figure \ref{fig:Direct} and for cycle 26 using wavelet method on the relationship between F10 and $\delta\nu$   (see Figure \ref{fig:dnu_RF}).
Overall, we find from our analyses that the $p$-mode frequencies have already reached their maximum around the beginning of 2025 for solar cycle 25 and are now in the descending phase. Also, given our prediction of the SC~26 minimum near 2031, the descending phase of SC~25 is expected to last approximately seven years, which is similar to the seven‑year decay phase observed for SC~23. Clearly, SC25 is stronger than SC24 and is more comparable to SC23 although the small-scale features are different in the two.
 
Our analyses also warrants improvements in machine-learning based methods if we are to use them for data-driven, time series solar activity forecasts. A careful examination of the scaled-forecasts in our study show that Deep learning-NBEATS method has been unable to reproduce the QBOs like the other methods, despite all showing similar qualitative behaviour. We leave such improvements for future. 

The cyclic pattern of $p$-mode frequency shifts is similar to that of radio flux, sunspot numbers, and many other proxies \citep{Chaplin2007}. This suggests that solar activity proxies are influenced by different layers of the Sun from the sub-photospheric layers up to the corona. The scaled-forecasts of $p$-mode frequency shifts based on their relationship with activity proxies such as radio flux and sunspot numbers, exploited in this study, are thus expected to be sensitive to the type and extent of data. Different type of dataset have different sources and magnitudes of errors as they are measured by different instruments and sample different regions of the Sun. Also, very small-scale fluctuations get smoothed out with data points averaged over 9 days. We have therefore, not shown any error bars for the forecasts and no quantitative conclusion on magnitudes are emphasised here.  We conclude that future improvements in forecasting model and with more data, there is potential for $p$-mode frequency shifts to act as an additional and independent indicator of solar activity and which will provide a valuable link from inner regions of the Sun to the outer atmosphere for space weather forecasting.

Although the first attempts to characterise $p$-mode frequency shifts during SC25, made in this article using either the past measured frequency shifts or their relationship with SSN and F10, is promising, its direct interpretability is limited at this stage. Our knowledge of the physical processes inside the Sun comes from many other helioseismic parameters such as amplitudes, line-widths etc. Also, there are many other solar activity indices that show similar variation with solar cycle as $p$-mode properties. Including these parameters and indices along with longer dataset of frequencies will be essential in establishing more accurate and more detailed machine-learning-based forecast methods.

We conclude that the $\delta\nu$ forecast for the remaining period of SC25 computed here, directly from the measured dataset of $\delta\nu$ and complemented by the scaled-frequency shifts (exploiting relationship of $\delta\nu$ with SSN and F10), using machine-learning methods provided an alternative way to study future evolution of oscillation frequencies with solar activity. However, helioseismic monitoring of the Sun needs to be continued for the foreseeable future so that robust forecasting methods can be developed further to understand solar activity and its evolution through  $p$-mode diagnostics.  %This new capability 
The results presented here represent a nascent step towards using  oscillation frequencies as an additional and independent indicator of solar activity. 

\begin{acks}

RJ and AK acknowledge partial funding by the STFC (UK) Impact Acceleration Award via The University of Sheffield's Internal Knowledge Exchange Scheme. The code was developed by AK using the Google Colab cloud computing platform. For analyses and visualization, open-source Python packages NumPy, Pandas, PyWT, Matplotlib, LightGBM, and N-BEATS were used. We also acknowledge the use of Gemini (Google) and ChatGPT (OpenAI) for occasional code debugging and suggesting edits for improvement. This work also utilizes GONG data obtained by the NSO Integrated Synoptic Program, managed by the National Solar Observatory, which is operated by the Association of Universities for Research in Astronomy (AURA), Inc. under a cooperative agreement with the National Science Foundation (USA) and with contribution from the National Oceanic and Atmospheric Administration (NOAA). The GONG network of instruments is hosted by the Big Bear Solar Observatory, High Altitude Observatory, Learmonth Solar Observatory, Udaipur Solar Observatory, Instituto de Astrof\'{\i}sica de Canarias, and Cerro Tololo Interamerican Observatory. 
\end{acks}
%\section{Additional statments}

\begin{authorcontribution} RJ wrote the main manuscript text with contributions from the co-authors. The code
was developed by AK using the Google Colab cloud computing platform. The frequencies corresponding to
9-days were generated for this study by the co-author SCT using the GONG project pipeline. All authors
reviewed the manuscript.
\end{authorcontribution}

\begin{dataavailability}
The Sunspot dataset used in this study is publicly available at \url{https:www.sidc.be/silso/datafiles}. The 10.7 cm Radio flux data can be obtained from \url{https://www.spaceweather.gc.ca} The frequencies corresponding to 9‑days were generated for this study by the co‑author SCT using the GONG project pipeline. 
\end{dataavailability}

%\begin{fundinginformation}
%RJ and AK acknowledge partial funding by the STFC (UK) Impact Acceleration Award via The University of Sheffield's Internal Knowledge Exchange Scheme.
%\end{fundinginformation}
%\begin{ethics}
%\begin{CompetingInterests} The authors declare no competing interests.
%\end{CompetingInterests}
%\end{ethics}

\bibliographystyle{spr-mp-sola}
\bibliography{main.bib}
\end{article}
\end{document}